\documentclass[a4paper,aps,reprint,pra, 11pt, onecolumn, superscriptaddress,nofootinbib, showpacs, notitlepage]{revtex4-1} 
\usepackage{times}
\usepackage{array}
\usepackage{amsmath}
\usepackage{amsfonts}
\usepackage{amssymb}
\usepackage{amsthm}
\usepackage{amsbsy}
\usepackage{stmaryrd}
\usepackage{pifont}
\usepackage{latexsym}
\usepackage{mathtools}
\usepackage{bbold}
\usepackage{gensymb}
\usepackage[all]{xy}
\usepackage{graphicx}
\usepackage{braket}

\usepackage{epstopdf}
\usepackage{epsfig}
\usepackage{verbatim}
\usepackage{multirow}
\usepackage{rotating}
\usepackage{minibox}
\usepackage{tabularx}

\usepackage[titletoc,toc,title]{appendix}
\usepackage{enumerate}
\usepackage{enumitem}
\usepackage{color}
\usepackage[dvipsnames]{xcolor}
\usepackage[right=2.75cm, left=2.75cm, top=3cm, bottom=3cm]{geometry}

\usepackage{comment}
\usepackage[colorlinks=True, linkcolor=Red,citecolor=Green]{hyperref}
\usepackage{sidecap}
\usepackage{bbm}
\usepackage{bm}

\begin{document}

\title{Quantum superposition of spacetimes obeys Einstein's Equivalence Principle}

\author{Flaminia Giacomini}%
 \email{fgiacomini@perimeterinstitute.ca}
\affiliation{%
Perimeter Institute for Theoretical Physics, 31 Caroline St. N, Waterloo, Ontario, N2L 2Y5, Canada
}%
\author{\v{C}aslav Brukner}
\affiliation{%
	Vienna Center for Quantum Science and Technology (VCQ), Faculty of Physics, University of Vienna, Boltzmanngasse 5, A-1090 Vienna, Austria
}%
\affiliation{%
	Institute of Quantum Optics and Quantum Information (IQOQI), Austrian Academy of Sciences, Boltzmanngasse 3, A-1090 Vienna, Austria
}%

\begin{abstract}
	We challenge the view that there is a basic conflict between the fundamental principles of Quantum Theory and General Relativity, and in particular the fact that a superposition of massive bodies would lead to a violation of the Equivalence Principle. It has been argued that this violation implies that such a superposition must inevitably spontaneously collapse (like in the Di{\'o}si-Penrose model). We identify the origin of such an assertion in the impossibility of finding a local, classical reference frame in which Einstein's Equivalence Principle would hold. In contrast, we argue that the formulation of the Equivalence Principle can be generalised so that it holds for reference frames that are associated to quantum systems in a superposition of spacetimes. The core of this new formulation is the introduction of a quantum diffeomorphism to such Quantum Reference Frames (QRFs). This procedure reconciles the principle of linear superposition in Quantum Theory with the principle of general covariance and the Equivalence Principle of General Relativity. Hence, it is not necessary to invoke a gravity-induced spontaneous state reduction when a massive body is prepared in a spatial superposition.
\end{abstract}

\maketitle

\emph{Article prepared for the Special Topic Collection ``Celebrating Sir Roger Penrose's Nobel Prize''}

\vspace{0.5cm}

It is a widespread view that the fundamental principles of General Relativity and Quantum Theory are incompatible. These two theories are famously hard to combine, partially because their unification presents technically hard challenges, but also because it is believed that in order to find a theory of Quantum Gravity the very notions of space, time, and the fundamental principles of these two theories should undergo a deep modification. Because of the difficulties involved in this process, some authors have wondered whether it is necessary to find a quantum description of the gravitational field at all~\cite{diosi1987universal, diosi1989models, penrose1996gravity, Carlip:2008zf, Bassi:2017szd}. While the commonly held view is that we should quantise gravity, according to these authors we should adapt the laws of Quantum Theory to the classical description of the gravitational field.

Roger Penrose argued in Refs.~\cite{penrose1996gravity, penrose2014gravitization} against the possibility of keeping a massive body which can curve the spacetime in a quantum superposition of different positions. According to him, the root of the problem in realising such a spatial superposition of a massive body lies in the incompatibility between the fundamental principles of General Relativity and Quantum Theory: on the one hand, the principle of general covariance and the Equivalence Principle, and on the other hand, the principle of linear superposition. Modifications and violations of the Equivalence Principle and geodesic motion to account for quantum effects have been studied in the literature, e.g., in Refs~\cite{greenberger1970theory, greenberger1970theory2, aharonov1973quantum, lammerzahl1996equivalence, viola1997testing, rosi2017quantum, zych2018quantum, anastopoulos2018equivalence,seveso2017does, hardy2020implementation, pipa2019entanglement}.

\begin{figure}[t]
	\begin{center} 
		\includegraphics[scale=0.6
		]{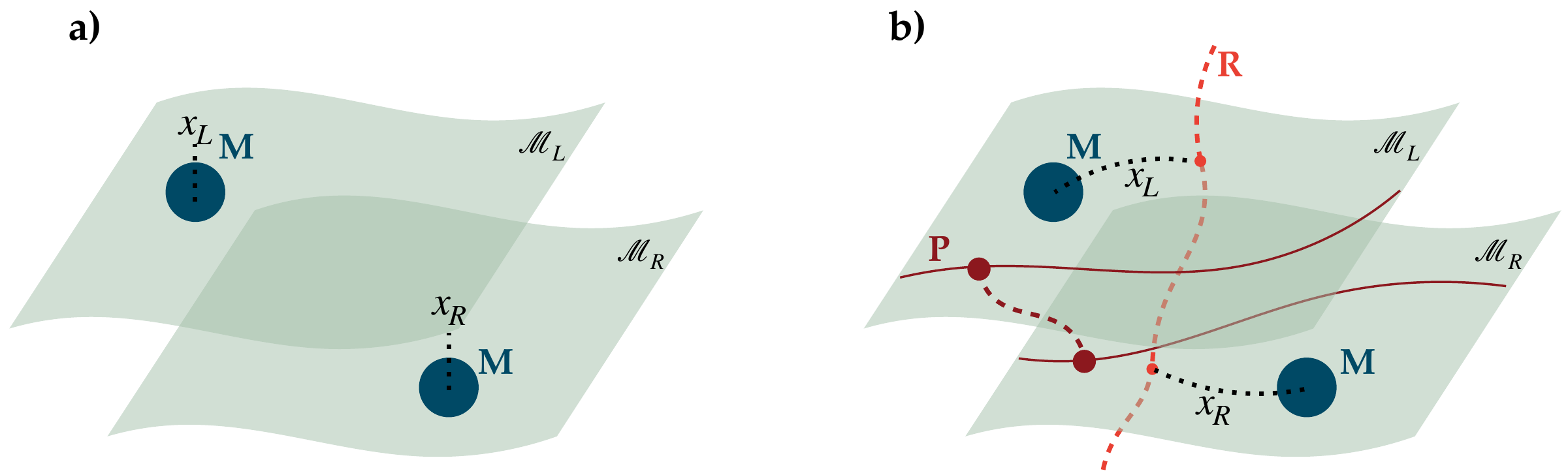}
		\caption{\label{fig:supmass} A massive system $M$ in a spatial superposition according to the description given by Penrose, in \textbf{a)}, and using the perspective of Quantum Reference Frames (QRFs), in \textbf{b)}. To each classical configuration of the mass there corresponds a classical manifold. In particular, we associate the manifold $\mathcal{M}_L$ to the $\ket{x_L}_M$ position of the mass, and the manifold $\mathcal{M}_R$ to the $\ket{x_R}_M$ position of the mass. To each manifold we associate a classical gravitational field, respectively $\ket{G_L}$ and $\ket{G_R}$, which can be considered as coherent states (the detailed description of this quantum state plays no role for the argument presented here).\\
		 \textbf{a)} Penrose describes this situation using an abstract labelling of points in the manifold. In our view, in this description there is a discrepancy between the fact that the labelling of points is not physical (which would imply that physics is translation invariant), and the fact that the two configurations of the gravitational field are taken to be distinguishable. Furthermore, it is not possible to identify points belonging to different spacetimes.\\
		 \textbf{b)} Using the tools of QRFs, the labelling of points is not abstract, but corresponds to the distance between two physical systems, in this case between the massive particle $M$ and the origin of the QRF, associated to the location of the initial QRF $R$. The initial QRF is decoupled from the rest of the physical systems, and defines the origin of the coordinate system. We can generalise the Equivalence Principle by changing to the QRF of any freely-falling particle $P$. Such a particle $P$ lives in the same manifolds $\mathcal{M}_L$ and $\mathcal{M}_R$, and its state is to be considered entangled with the mass $M$ and the gravitational field. More precisely, in each manifold the particle evolves according to the dynamical equation for a quantum system in a curved spacetime relative to the manifold. The classical trajectory in each manifold is illustrated by the red solid lines in the picture. A pair of points along the two geodesic trajectories can be operationally identified in terms of the distance between particle $P$ and the initial QRF $R$. In the new QRF, the distance between two objects is again physically meaningful, and the identification of points in each amplitude of the mass is to be performed by making use of this notion of distance between two physical systems, i.e., the mass $M$ and the probe particle $P$, which is at the origin of the new QRF. In our view, including explicitly the state of the QRF is a necessary prerequisite in order to meaningfully write the quantum superposition, which otherwise has no physical meaning.}
	\end{center}
\end{figure}

Here, we argue that a generalised formulation of the Einstein Equivalence Principle, obtained in Ref.~\cite{giacomini2020einstein} by using tools from the formalism of Quantum Reference Frames (QRFs), allows us to reconcile these fundamental principles. As a consequence, the spatial superposition of massive bodies is not to be considered as giving rise to inconsistencies between General Relativity and Quantum Theory. The main differences between Penrose's description and ours is illustrated in Fig.~\ref{fig:supmass}.

QRFs have been studied since the sixties in the context of both quantum information theory~\cite{aharonov1} and quantum gravity~\cite{dewitt1967quantum}. In quantum information, QRFs have played an important role in describing communication protocols in the absence of a shared reference frame between the emitter and the receiver, and in terms of resources, as bounded reference frames, see e.g. Refs.~\cite{aharonov1, aharonov2, aharonov3, brs_review, spekkens_resource, kitaev_superselection, palmer_changing, bartlett_degradation, smith_quantumrf, poulin_dynamics, skotiniotis_frameness, poulin_reformulation, poulin_deformed, busch_relational_1, busch_relational_2, busch_relational_3, jacques, angelo_1, angelo_2, angelo_3}. In quantum gravity, they have been discussed in the context of spacetime relationalism~\cite{rovelli_quantum}. Here, we take a different view of QRFs, which was introduced in Ref.~\cite{QRF} and further developed in Refs.~\cite{perspective1, perspective2, giacomini2019relativistic, yang2020switching, de2020quantum, streiter2021relativistic, krumm2020, ballesteros2020group, giacomini2021spacetime}. In this view, we associate a QRF to the state of an arbitrary quantum system, and introduce a method to transform to a different QRF, which can be in a quantum superposition or entangled state from the point of view of the initial QRF.

We start by reviewing Penrose's argument. We do not wish to give a complete account of Penrose's view, instead we will only highlight the elements which we will need in our analysis. In order to illustrate the argument, let us first consider the situation when a quantum system $M$ is in a quantum superposition state of two positions and all gravitational effects are ignored (i.e., we consider Minkowski spacetime). This situation is invariant under global translations. We prepare the quantum state in a coherent superposition of two stationary states (i.e., eigenstates of the Hamiltonian) $\ket{\psi}_M$ and $\ket{\chi}_M$. The two states are related by a translation, i.e., $\ket{\psi}_M = e^{-\frac{i}{\hbar}d \hat{P}} \ket{\chi}_M$, where $d$ is the distance between the centre of the two wavepackets, and $\hat{P}$ is the generator of translations. If coordinates have no physical meaning, then we should not be able to distinguish the two states corresponding to the alternative locations of the mass. If the dynamics is governed by a translationally invariant Hamiltonian $\hat{H}$, the two amplitudes are eigenstates with the same energy eigenvalue, i.e., $\hat{H}\ket{\psi}_M = v \ket{\psi}_M$ and $\hat{H}\ket{\chi}_M = v \ket{\chi}_M$. Any linear combination
\begin{equation} \label{eq:Minksup}
	\lambda \ket{\psi}_M + \mu \ket{\chi}_M
\end{equation}
is in turn an eigenstate with the same energy eigenvalue (departing from Penrose's line of argument, we note that, operationally, if the two vectors represent two equivalent quantum states, they are not distinguishable, and hence any resulting superposition is also identical to the quantum state). Hence, the superposed state is also a stationary state. In Penrose's analysis, this does not contradict the principle of general covariance, because all the various states of location of the system are physically equivalent and no relative phase is accumulated when the state is dynamically evolved.

Let us now take into account the gravitational field associated to the configuration of the quantum system. In this case, we have to modify the full state as
\begin{equation} \label{eq:Gensup}
	\lambda \ket{\psi}_M \ket{G_{\psi}}_G + \mu \ket{\chi}_M \ket{G_{\chi}}_G,
\end{equation}
where we can think of the states of the gravitational field $G$ as two coherent classical-like states. The exact form of the quantum states of the gravitational field does not matter for the purpose of the following analysis, but we take the two geometries to differ significantly from each other (in Quantum Theory, if two states are (macroscopically) distinguishable, they are orthogonal). According to the principle of general covariance, the labelling of points has no physical meaning. In General Relativity, one identifies physical points in spacetime based on coincidences between fields. Here, however, we consider no fields other than gravitational fields associated to two configurations of a massive system, and hence have no way of identifying physical points.
To overcome this problem, Penrose proposed to adopt an approximate identification of points in different spacetimes, for instance, in the Newtonian limit. However, even in this limit, when the metric is not well-defined, there is no \emph{single} reference frame transformation which makes the spacetime locally Minkowskian along a geodesic. As a result, we incur into a violation of the Equivalence Principle, i.e., the violation of the universality of free-fall. In this situation, each classical configuration of the gravitational field has a different timelike Killing vector, which identifies the direction of a time-translation. The argument then states that because of the relation between timelike Killing vectors and energy conservation laws, when the former are not well-defined the global state is not stationary. The difference in the energies results in an energy-time uncertainty principle, leading to the instability of the superposed quantum state. Furthermore, according to Refs.~\cite{penrose1996gravity, penrose2014gravitization}, the energy difference between the two amplitudes would lead to a quantum state which is a superposition of different vacua of the gravitational field, each of which would have a different notion of positive energy.

Crucially, the instability of the quantum state derives from the superposition of spacetime geometries. According to Penrose's argument, the validity of the Equivalence Principle is restored via a gravitationally-induced spontaneous state reduction~\cite{penrose1996gravity, penrose2014gravitization, howl2019exploring}: the quantum superposition of masses, initially in an unstable configuration, evolves into a classical, well-defined spacetime in some time $t_\Delta \approx \frac{\hbar}{E_\Delta}$, with $E_\Delta$ being the gravitational self-energy of the difference between the mass distributions of the two mass configurations. However, in Quantum Theory, the principle of linear superposition can in principle be applied to every massive system, as there is no built-in mass scale in the theory setting an upper bound to its applicability.

Penrose's argument rests on the (implicit) assumption that all coordinate systems are in a classical relation to each other. With this assumption, it is impossible to find a classical, local coordinate system in which the metric can be made locally minkowskian for both configurations of the mass. If we instead take the view, already advocated by Einstein, that coordinate systems correspond to physical rods and clocks, then we may reason as follows
\begin{enumerate}
	\item Physical systems are ultimately quantum, and hence can be in a superposition, or entangled, with other physical systems;
	\item All reference frames, including those associated to quantum systems, provide an equally good perspective from which physical laws can be described.
\end{enumerate}
Notice that, in this view, the position of a system in a specific choice of coordinates is not an abstract labelling of a manifold, but is defined as a distance between two physical systems, one of which serves as the origin of the coordinate system. We have introduced the notion of a Quantum Reference Frame (QRF)~\cite{QRF, perspective1, perspective2, giacomini2019relativistic, yang2020switching, de2020quantum, streiter2021relativistic, krumm2020, ballesteros2020group, giacomini2020einstein, giacomini2021spacetime}, i.e., a reference frame associated to a quantum system, whose state can be in a superposition or entangled with other physical systems. The key feature of the formalism that we exploit here is the generalisation of the standard diffeomorphism to a quantum superposition of classical diffeomorphisms. In the following, we follow the results of Ref.~\cite{giacomini2020einstein} and show that such a generalised coordinate transformation allows us to find a local coordinate system associated to a quantum particle, in which the metric field is locally minkowskian at the origin of the QRF. We call such a QRF a \emph{Quantum Locally Inertial Frame} (QLIF). In what follows, we are interested in QRFs defining relative coordinates between two quantum systems, however the formulation is more general.

\begin{figure}[t]
	\begin{center} 
		\includegraphics[scale=0.7
		]{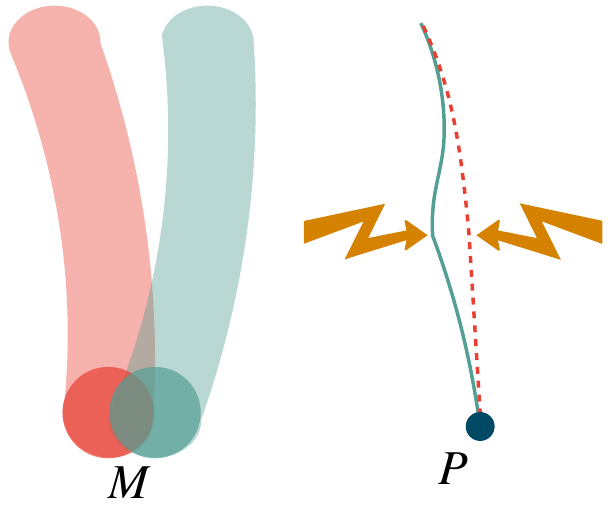}
		\caption{\label{fig:violationEEP} A mass $M$ is prepared in a spatial quantum superposition. A test particle $P$ is in the superposition of gravitational fields sourced by the mass $M$. According to Penrose, the situation results in a violaton of the Equivalence Principle because there is no classical reference frame in which the metric is locally minkowskian (i.e. there is no freely falling reference frame whose geodesic is a well-defined trajectory). The application of the principle of linear superposition to this situation would lead the particle $P$ to become entangled with the mass $M$ and its gravitational field. In our generalised Equivalence Principle, this situation is exactly what one should expect: for each position of the mass there is one geodesic trajectory for the freely falling particle $P$. It would only be possible to obtain a standard geodesic equation if we applied a non-inertial force (which we represent in the figure with the orange arrows) to the particle. Hence, the reference frame associated to the particle would not be inertial. }
	\end{center}
\end{figure}

Using the tools of QRFs, we can reach a different conclusion to Penrose, which does not require us to ``gravitise'' Quantum Theory (in his wording). Instead, we show that the principle of general covariance and the Equivalence Principle can be generalised to the quantum framework. These principle, in their generalised formulation, are compatible with the principle of linear superposition. Hence, a mass in a spatial superposition is not to be considered as an unstable configuration, but it is a perfectly legitimate physical situation. As shown in Ref.~\cite{giacomini2020einstein}, a test particle $P$ falling freely in a superposition of spacetimes sourced by a mass $M$ in an initial QRF $R$ becomes entangled with $M$ in such a way that in each amplitude it follows a classical geodesic relative to the metric. The Equivalence Principle is obeyed in each amplitude and hence in the superposition. In contrast, a test particle following any arbitrary (definite or classical) standard trajectory would violate the Equivalence Principle, as illustrated in Fig.~\ref{fig:violationEEP}. Importantly, our results are derived in the same regime as in Penrose's work, i.e., i) each state corresponds to a classical configuration of the gravitational field, ii) the states of the gravitational field are macroscopically distinguishable (and hence orthogonal to each other), and iii) the states can be linearly superposed. 

Our construction automatically implements the fact that absolute coordinates on a manifold have no physical meaning, and only defines the physically relevant relative coordinates. Such relative coordinates correspond to the distance between any physical system and the initial QRF $R$, defining the origin of a coordinate system. This allows us to operationally identify points belonging to different spacetimes, and meaningfully write the superposition state by adopting a relational view: the QRF $R$ is always at the origin of the coordinate system, and the quantum states of $M$ and $P$ are the relative states, in position basis, of $M$ and $P$ as seen from $R$. In this view, there is no contradiction in having different Killing vectors which are in a superposition in each amplitude. Instead, when a quantum system is the source of the gravitational field, and hence the spacetime is not classical, new interesting effects arise, for instance that the time shown by a quantum clock runs in a quantum superposition or not according to the QRF chosen to describe it~\cite{castro2020quantum}. Nonetheless, even when this is the case from the point of view of one QRF, one can always find another QRF in which it flows at a well-defined rate. This has already been shown in different contexts~\cite{castro2020quantum, giacomini2021spacetime}. The prediction is compatible with a notion of extended covariance of physical laws~\cite{QRF} and with a generalised group of symmetries, so far explicitly derived in the Galilean context~\cite{ballesteros2020group}. In Quantum Field Theory, superpositions of different modes leading to different vacua of the theory has been shown to be consistent in the case of the Unruh effect in Refs.~\cite{Barbado:2020snx, Foo:2020xqn}.

We can now formulate our extension of the Einstein Equivalence Principle, first introduced in Ref.~\cite{giacomini2020einstein}. We show that there always exists a unitary QRF transformation such that the gravitational field, and the superposition of gravitational fields, can be made locally minkowskian in the QRF associated to the position of any quantum particle $P$, thus generalising the notion of Locally Inertial Frames (LIFs) to Quantum Locally Inertial Frames (QLIFs). We then formulate the following generalisation of the Einstein's Equivalence Principle
\begin{verse}
	\emph{In any and every \textbf{Quantum} Locally Inertial Frame (QLIF), anywhere and anytime in the universe, all the (nongravitational) laws of physics must take on their familiar non-relativistic form.} 
\end{verse}
The most important element of this generalisation is the construction of a QRF transformation to the QLIF of the particle serving as the QRF. Crucially, we find that this transformation is unitary even in the superposition of spacetimes that we have previously discussed. Specifically, we start from the initial QRF $R$, from which we describe a quantum particle $P$ and the state of the gravitational field. Following Ref.~\cite{giacomini2020einstein}, to which we refer the interested reader for details, from the perspective of $R$ the joint state of the massive system $S$, the local gravitational field at the location of $P$, and the particle $P$, is
\begin{equation} \label{eq:PsiR}
	\ket{\Psi}^{(R)}= \frac{1}{2\sqrt{2}} \sum_{i=L,R} \int d^4 x \sqrt{-g_i(x)} \psi_i(x) \ket{i}_S\ket{g^i(x)} \ket{x^{(i)}}_P \ket{0}_R, 
\end{equation}
where $\ket{i}_S$, with $i=L, R$, is the quantum state of the massive system $S$ in position basis, $\ket{g^i(x)}$ is the classical configuration of the gravitational field at the distance $x$ from the origin of the QRF $R$, endowed with the scalar product
\begin{equation}
	\frac{1}{4}\braket{g_i(x)| g_j(x')}\braket{x^{(i)}|x'^{(j)}}_P = \frac{\delta^{(4)}(x-x')}{\sqrt{-g_i(x)}}\delta_{ij},
\end{equation}
 with $\sqrt{-g_i(x)}= \sqrt{|\det g_i(x)|}$, and $\psi_i(x)$ is the wavefunction in position basis of the relative state of $P$ as seen from $R$. In Eq.~\eqref{eq:PsiR}, the state of the initial QRF $\ket{0}_R$ is only an auxiliary state, and does not correspond to a physical degree of freedom. Notice that $P$ is entangled with the rest of the quantum systems. We now build the transformation to the QLIF of particle $P$ by linearly superposing a standard transformation to a LIF in General Relativity for each value of the position of $P$, labelled by $x$, and each configuration $i$ of the gravitational field. In order to transform to a LIF, it is enough, in General Relativity, to perform a linear transformation from a coordinate system labelled with $x'$ to a coordinate system labelled with $\xi$ of the type
\begin{equation}
	\xi^\mu = b^\mu_\alpha (x'-x)^\alpha, \qquad x'^\alpha =x^\alpha +  f_\mu^\alpha \xi^\mu,
\end{equation}
where the coefficients $b^\mu_\alpha = b^\mu_\alpha (x, g_i)$ and $f_\mu^\alpha = f_\mu^\alpha (x, g_i)$ are functions of the point $x$ about which we expand the transformation and of the metric field $g_i$. In addition, we have the property that, to the leading order of approximation, $f_\mu^\alpha b^\mu_\beta = \delta^\alpha_\beta$ and $b^\mu_\alpha f_\nu^\alpha  = \delta^\mu_\nu$.
 In Quantum Theory, this transformation is represented as a unitary operator controlled on the position of $P$ and on the index $i$ (see the explicit form of the unitary transformation in Ref.~\cite{giacomini2020einstein}). After the transformation, and evaluating the field at the origin of the final QRF $P$, i.e., for $\xi=0$, we find that the gravitational field is minkowskian at the origin of the QRF
 \begin{equation}
 	\ket{\tilde{\Psi}}^{(P)} = \frac{1}{2\sqrt{2}}\sum_{i=L,R} \int d^4 x \sqrt{-g_i(x)} \psi_i(x) \ket{-x^{(i)}}_R \ket{\xi_i}_S \ket{\eta} \ket{0^{(i)}}_P,
 \end{equation}
 where $\xi_i^\mu = b^\mu_\alpha i^\alpha$ and $\eta$ is the Minkowski metric. Hence, we have shown that it is possible to build a unitary QRF transformation to a QLIF such that the metric field can be made locally minkowskian at the origin of the QRF, also in the case of a superposition of spacetimes.

In this paper, we have reviewed how Quantum Reference Frames allow us to generalise the Einstein's Equivalence Principle~\cite{giacomini2020einstein}, by constructing a transformation to a Quantum Locally Inertial Frame. Thanks to this transformation, we can make the metric field locally minkowskian even in the presence of a superposition of classical spacetimes, arising when a massive body is prepared in a spatial superposition. Hence, our argument overcomes the criticism, initially brought forward by Roger Penrose, that the superposition of massive bodies only has a finite lifetime because it violates one of the fundamental principles of General Relativity. Penrose's argument is that rather than quantising gravity we should ``gravitise'' quantum theory. However, we have shown here that there is no fundamental incompatibility between the principles of these two theories, provided that we revise the formulation of the Equivalence Principle in such a way that it holds in all reference frames that quantum theory allows for. 

\acknowledgements{F.G. acknowledges support from Perimeter Institute for Theoretical Physics. Research at Perimeter Institute is supported in part by the Government of Canada through the Department of Innovation, Science and Economic Development and by the Province of Ontario through the Ministry of Colleges and Universities. \v{C}.B. acknowledges financial support from the Austrian Science Fund (FWF) through BeyondC (F7103-N48), the European Commission via Testing the Large-Scale Limit of Quantum Mechanics (TEQ) (No. 766900) project, the Foundational Questions Institute (FQXi) and the Austrian-Serbian bilateral scientific cooperation no. 451-03-02141/2017-09/02. This publication was made possible through the support of the ID 61466 grant from the John Templeton Foundation, as part of The Quantum Information Structure of Spacetime (QISS) Project (qiss.fr). The opinions expressed in this publication are those of the author(s) and do not necessarily reflect the views of the John Templeton Foundation.}

\subsection*{Data Availability}
Data sharing is not applicable to this article as no new data were created or analyzed in this study.

\subsection*{Conflict of interest}
The authors have no conflicts to disclose.

\section*{References}
\bibliography{biblioPenrose}{}
\bibliographystyle{ieeetr}
\end{document}